\documentclass[manuscript]{aastex}

\slugcomment{submitted to ApJ}

\shorttitle{HC$_3$N J=5--4 in Egg nebula}
\shortauthors{Dinh-V-Trung \& J. Lim}

\begin{document}


\title{The shaping effect of collimated fast outflows in the Egg nebula}

\author{Dinh-V-Trung\footnote{on leave from Center for Quantum Electronics, Institute of Physics,
Vietnamese Academy of Science and Technology, 10 DaoTan, ThuLe, BaDinh, Hanoi, Vietnam}, Jeremy Lim}
\affil{Institute of Astronomy and Astrophysics, Academia Sinica, Taiwan}
\email{trung@asiaa.sinica.edu.tw, jlim@asiaa.sinica.edu.tw}

\begin{abstract}
We present high angular resolution observations of the HC$_{\rm 3}$N J=5--4 line
from the Egg nebula, which is the archetype of protoplanetary nebulae. 
We find that the HC$_{\rm 3}$N emission in the approaching and receding portion of the envelope traces a 
clumpy hollow shell, similar to that seen in normal carbon rich envelopes. 
Near the systemic velocity,
the hollow shell is fragmented into several large blobs or arcs with missing portions 
correspond spatially to locations of previously reported high--velocity outlows in the
Egg nebula. This provides direct evidence for the disruption of the slowly--expanding  
envelope ejected during the AGB phase by the 
collimated fast outflows initiated during the transition to the protoplanetary nebula phase.
We also find that the intersection of fast molecular outflows 
previously suggested as the location of the central post-AGB star is significantly offset from the
center of the hollow shell. 
From modelling the HC$_3$N distribution we could reproduce
qualitatively the spatial kinematics of the HC$_3$N J=5--4 emission
using a HC$_3$N shell with two pairs of cavities cleared by the collimated high velocity
outflows along the polar direction and in the equatorial plane.
We infer a relatively high abundance of HC$_3$N/H$_2$ 
$\sim$3x10$^{-6}$ for an estimated mass--loss rate of 3x10$^{-5}$ M$_\odot$ yr$^{-1}$ in the
HC$_3$N shell.
The high abundance of HC$_3$N and the presence of some weaker J=5--4 emission in the
vicinity of the central post-AGB star suggest an unusually efficient formation of this molecule in the Egg nebula. 
\end{abstract}

\section{INTRODUCTION}
The rapid evolution of low and intermediate mass stars (1 to 10 M$_\odot$) after the end of the Asymptotic 
Giant Branch (AGB) phase is accompagnied by a radical change in the morphology of their circumstellar envelopes
around them. The circumstellar envelope created by the slow and dusty stellar wind during the AGB phase is
known to be roughly spherically symmetric as the radiation pressure on dust grains is expected
to be isotropic. On the other hand, a variety of morphologies ranging from spherical to multipolar shapes have been
observed in post-AGB envelopes and planetary nebulae. 
It is also during this phase that collimated high velocity outflows are seen, 
such as in the Egg nebula (Cox et al. 2000) and in CRL 618
(S\'{a}nchez-Contreras et al. 2004). The shaping of the complex envelope morphology and the mechanism 
that generates the high--velocity
outflows of post-AGB stars are very poorly understood. Lee \& Sahai (2003) suggest that envelopes with bipolar morphologies are
shaped by the interactions with a fast collimated
outflow or jet launched in the polar directions. Their hydrodynamic
simulations are able to reproduce both the envelope morphology and other properites such intensity of emission lines
excited in the shocked gas at the interface between the jet and the slow wind.

The Egg nebula (CRL 2688) is widely considered as the prototype of proto-planetary nebulae (PPN),
a class of stars in the rapid transition phase between the AGB
and planetary nebulae. During the PPN phase the central post-AGB star contracts and gradually becomes
hotter, but is not yet hot enough to ionize its surrounding envelope. Using
high resolution spectroscopic observations of the scattered stellar light
Klokova et al. (1996) determine a spectral type F5Ia
for the central star with an effective temperature of T$_{\rm eff}$=6500K. 
The abundance analysis reveals the enrichment of C and N together with
strong enhancement of slow neutron capture (s-process) elements, as expected for the
carbon rich post-AGB star at the center of the Egg nebula.

Recently, using multi-epoch optical images from Hubble Space Telescope, Ueta et al. (2006)
succesfully measured the spatial expansion of the nebula and thereby  
determined the distance to CRL 2688 of 420 pc. We shall henceforth use 
this distance for the Egg nebula

High spatial resolution optical images of CRL 2688 reveal 
a pair of bipolar lobes seen as search-light beams emanating
from the central (obscured) star and oriented perpendicular to a dark lane (Sahai et al. 1998a). 
This conspicuous dark lane has been commonly
interpreted as an equatorial disk of cold dust. High angular resolution images of hot 
molecular hydrogen gas emission at 2.2 $\mu$m reveal shocked gas, 
tracing the strong interaction between the fast outflows and
the slow wind (Sahai et al. 1998b). Surprisingly, hot molecular hydrogen emission is seen in both bipolar lobes
as well as in the equatorial plane far beyond the dark lane, indicating the presence of 
multiple fast collimated outflows.

Indeed, high spatial resolution mapping of CO J=2--1 emission by Cox et al. (2000) reveals the 
presence of several pairs of collimated fast molecular outflows along both the bipolar axis and
the equatorial plane. These molecular outflows can be traced back to a common origin, presumably the
location of the central post-AGB star in the Egg nebula. 
Cox et al. (2000) suggest that
the observed CO emission does not constitute the actual fast collimated outflows, but rather
molecular gas swept up by even faster collimated outflows that comprise mainly 
atomic or ionized gas. By contrast to its appearance in the optical, no equatorial disk is the
Egg nebula is seen in the CO J=2--1 by Cox et al. (2000). 
High spatial resolution observation of HCN J=1--0 
by Bieging \& Nguyen-Q-Rieu (1996) traces molecular gas at higher densities along both 
the bipolar lobes and the equatorial plane consistent with
gas swept up gas by the fast outflows. The only observation
that seems to directly trace the AGB wind is that by Yamamura et al. (1996) in $^{13}$CO J=1--0, which does
not exhibit any high velocity wings. The limited
spatial resolution (about 5\arcsec) of the observation, however, does not provide any detailed information
on the structure of the circumstellar envelope. 

Emision lines from cyanopolyyne molecules such as HC$_3$N are very prominent in the centimeter and
millimeter wavelengths region toward the Egg nebula. Cyanopolyyne molecules, as large as HC$_9$N, have been
detected (Truong-Bach et al. 1993). The line profiles of cyanopolyyne molecules do not 
exhibit high velocity wings, suggesting
that their emission originate from the slowly expanding shell ejected during the AGB phase.
Current chemical models (Millar et al. 2000, Cherchneff et al. 1993) for carbon rich envelopes suggest that 
HC$_3$N molecules are formed by photochemistry in the outer part of the expanding envelope. Thus the
spatial distribution of HC$_3$N is predicted to exhibit a hollow shell structure. In
the case of the carbon rich envelope of IRC+10216, the spatial distribution of cyanopolyyne molecules 
has been imaged at high angular resolution (Bieging \& Tafalla 1993, Lucas \& Gu\'{e}lin 1999, 
Dinh-V-Trung \& Lim 2008) and shown to have the hollow shell structure as expected
from chemical models. The relatively high anbundance and high electric diople moment of HC$_3$N
make its rotational lines relatively strong in the circumstellar envelope. The HC$_3$N lines 
might be useful to probe not just the chemistry but also the structure
of the envelope at high angular resolution. In the Egg nebula we also expect the molecule to form
in the outer part of the envelope and its emission lines could be used to 
trace the spatial structure and kinematics of the outer envelope of the Egg nebula. Interferometric
observation of HC$_3$N emission line in the 3 mm band by Nguyen-Q-Rieu \& Bieging (1990) indicates
a compact and centrally peaked emission, most likely due to the limited angular resolution of
about 8\arcsec. 

In this paper we present high resolution observations of the HC$_3$N J=5--4 emission line, 
which traces the outer molecular envelope of the Egg nebula. The envelope is found to be
disrupted by the passage of the fast collimated outflows along the polar direction and in the
equatorial plane. We also present a simple model of the envelope to
better understand its spatial kinematics.

\section{OBSERVATION}
We observed the Egg nebula on 2002 November 24 and 2003 March 03 using the Very Large
Array (VLA\footnote{The National Radio Astronomy Observatory is a facility of the National 
Science Foundation operated under a cooperative agreement by Associated Universities, Inc.}) 
in its C and D configuration. We pointed the telescope at
$\alpha_{j2000}$=21h02m18.8s, $\delta_{j2000}$=36d41m38.0s, which is very close to the
peak of the continuum emission observed previously by Cox et al. (2001) and Jura et al. (2001).
The rest frequency of the HC$_3$N J=5$-$4 line 
as compiled in the Lovas/NIST database (Lovas 2004) is 45.490316 GHz. To observe this line
we configured the VLA correlator in the 2AC mode with 64 channels spanning a bandwidth of 6.25 MHz, 
thus providing a velocity resolution of 0.65 kms$^{-1}$ per channel over a 
useful velocity range of $\sim$40 kms$^{-1}$. The total on-source integration time is about 1 hour in
D configuration and about 4 hours in C configuration.
We monitored the nearby quasar 21095+35330 every 5 minutes to correct for
the antennas gain variations caused primarily by atmospheric fluctuations. The stronger quasar 2253+161 was 
used to correct for the shape of the bandpass 
and its variation with time. The absolute flux scale of our observations was determined from observations of 
standard quasars 0137+331 and 0410+769.
 
We edited and calibrated the raw visibilities using the AIPS data reduction package. 
The calibrated visibitilies from different configurations were merged using the task DBCON,
and then Fourier transformed to form the DIRTY images. We employed the robust   
weighting together with tapering of the visibilities to obtain a satisfactory compromise between angular resolution and 
sensitivity to extended emission. The DIRTY images were deconvolved
using the clean algorithm IMAGR implemented in AIPS, providing a synthesized beam of 1\arcsec.3x1\arcsec.08 at a 
position angle of PA=2$^\circ$.35.
The rms noise level in our channel maps of HC$_3$N J=5$-$4 is 2.3 mJy beam$^{-1}$ in each 
velocity channel of 3.9 kms$^{-1}$. The conversion factor between the brightness 
temperature of the HC$_3$N J=5--4 emission and 
its flux density is $\sim$2.37 mJy K$^{-1}$. Table. 1 provides a summary of our observations.
\section{RESULTS}
Figure 1 shows our channel maps of the HC$_3$N J=5--4 emission. In Figure 2 we show the HC$_3$N J=5--4 line profile
derived by integrating the channel maps over a region where the emission is detected above 2$\sigma$ level. 
The HC$_3$N J=5--4 line has been previously observed by Fukasaku et al. (1994)
with the Nobeyama 45m telescope, which has a primary beam at FWHM of about 40\arcsec. 
Using the main beam efficiency provided in their paper, we estimate a conversion factor between main beam temperature
and flux density of 4 Jy/K, thus
giving a peak flux density of about 2 Jy for the HC$_3$N J=5--4 line in their observation. 
We measured in our observations (see Figure 2) a peak flux density for this line of $\sim$1.6 Jy.
The shape of the line profiles in both observations are also very similar, exhibiting a near 
parabolic shape
with a trough around the systemic velocity of --35 kms$^{-1}$. We conclude that
our VLA observation has recovered about 80\% of the emission in the HC$_3$N J=5--4 line present in the
abovementioned single--dish observation. 

In the channel maps the HC$_3$N J=5--4 line shows the typical characteristics of an 
expanding shell, i.e. the emission appears
largest near the systemic velocity, and contracts toward the center at progressively higher blueshifted and redsfhited
velocities. The emission spans velocities between --53.9 and --19.1 kms$^{-1}$, from which we estimate an
expansion velocity of $\sim$17 kms$^{-1}$, that is similar to that obtained by Truong-Bach et al. (1993).
At redshifted velocities between --26.9 to --19.1 kms$^{-1}$ and the blueshifted 
velocities between --53.9 to --46.2 kms$^{-1}$ the HC$_3$N J=5--4 emission traces a clumpy and hollow
shell--like structure with radius of $\sim$4\arcsec\, -- 5\arcsec. Such a hollow shell structure is expected
from the prediction of HC$_3$N formation in the envelopes of carbon rich stars 
(Cherchneff et al. 1993, Millar et al. 2000). We note that
similar structures have been seen most prominently in the prototypical carbon rich star IRC+10216 
(Bieging \& Tafalla 1993, Dinh-V-Trung \& Lim 2008). The large angular size of the hollow shell and
more importantly the lack of high velocity emission wings of the HC$_3$N J=5--4 line suggest that 
this line is tracing the remnant envelope created by the slow dusty wind from the central star of 
the Egg nebula during the AGB phase. Using the distance to the Egg nebula of 420 pc 
as estimated by Ueta et al. (2007), the linear size of the shell is about 3 10$^{16}$ cm.
We note that the radius of the HC$_3$N shell in the archetype carbon rich envelope
of IRC+10216, which is located at an estimated distance of 120 pc, is between 15\arcsec\, to 20\arcsec. 
If IRC+10216 was located at the same distance
as the Egg nebular, the HC$_3$N shell would have a radius of 4\arcsec\, to 6\arcsec.
Assuming that the interstellar radiation field is the same for both objects,
the similarity in size of the HC$_3$N shell in both the Egg nebula and IRC+10216 suggests that
the mass loss rate in both objects is comparable. The mass loss rate of IRC+10216 is estimated to be 
in the range 1 to 3 10$^{-5}$ M$_\odot$/yr from the modeling results of Crosas \& Menten (1997), 
Dinh-V-Trung \& Nguyen-Q-Rieu (2000) and Schoier et al. (2002). The comparably high mass loss rate 
of the central star in the Egg nebula at the time it produced the observed envelope suggests that
this period might correspond to the superwind phase the end stage of the evolution on AGB.  

Near the systemic velocity, most prominently at and between velocity channels --30.7 and --38.4 kms$^{-1}$, 
the hollow shell structure is clearly disrupted and fragmented into several large clumps 
of intense HC$_3$N J=5--4 emission. At the systemic velocity the shell is delineated by
four large clumps roughly mirror-symmetric with respect to the lines passing through the nebula center at
position angles of $\sim$20$^\circ$ and 110$^\circ$. The space between these clumps form two pairs
of cavities. Figure 3 shows the integrated intensity map of the HC$_3$N J=5--4 emission. 
This integrated intensity map clearly shows the presence of the cavity pair in the 
equatorial plane and also the pair along the polar direction. These cavities are 
separated by four intense peaks of HC$_3$N emission. These evacuated 
cavities are oriented along the axes of the fast collimated outflows as traced
by the molecular hydrogen emission (Sahai et al. 1998b) and CO J=2--1 emission (Cox et al. 2000) 
as sketched in Figure 3. As can be seen in the figure, the spatial correspondance between the cavities and 
the fast outflows is reasonably good.
These fast outflows are confined within the cavities. That is especially evident for the outflows
in the polar direction, which are closely aligned within the narrow cavities oriented at a position
angle of about 20$^\circ$. We note that Cox et al. (2000) esimate the average position angle 
of the polar outflows to be about 17$^\circ$, very similar to that of the cavities seen in our
HC$_3$N J=5--4 observations. Cox et al. (2000) also suggest that all the outflows can be traced
back to a common origin. As can be seen in Figure 3, the intersection of different outflows
is clearly offset from the center of the HC$_3$N shell. Close inspection of Figure 3 indicates that the shell
is centered close to the position of the 1.3 mm continuum peak ($\alpha_{j2000}$=21h02m18.647s, $\delta_{j2000}$=36d41m37.8s) 
observed by Cox et al. (2000). 
We estimate that the offset between the intersection of outflows and the center of the HC$_3$N shell is
about $\sim$1\arcsec, which is quite significant given our high angular
resolution of $\sim$1\arcsec.3. If the outflows indeed have a common source, that source might have moved in the envelope.
The dynamical age of the outflows estimated by Cox et al. (2000) ranges from 125 years to 1000 years.
Even if the source moves at a relatively slow speed of 2 kms$^{-1}$, its position could change by as much as 1\arcsec\,
in 1000 years.  

Interestingly, HC$_3$N J=5--4 emission is also
found close to center of the Egg nebula in several velocity channels between --30.7 to --42.3 kms$^{-1}$,
forming a faint bridge between the Northern and Southern portions of the shell.
The existence of HC$_3$N in the inner envelope is difficult
to understand within the framework of current chemical models.
Previous observations by Audinos et al. (1994) show that HC$_3$N emissions 
from higher lying transitions in the 2 mm and 1 mm band, which are excited mainly
under warmer and denser conditions, peak at the center of the 
carbon rich envelope IRC+10216, implying the much enhanced abundance of HC$_3$N toward
the center of the envelope. For the lower lying
transition J=5--4 to be seen toward the center of the envelope, the abundance of HC$_3$N needs 
to be even higher because of the lower optical depth in comparison to higher J transitions.
This suggests that the formation of  HC$_3$N is quite efficient in the inner dense region of the envelope 
where photodissociation of molecular species is low. There are two main differences in comparison
to the archetypical envelope of IRC+10216: the center of the Egg is likely quite disturbed because
of the passage of fast outflows and the central hotter (F-type) post-AGB star emits more UV photons.
These differences might contribute to the enhanced efficiency of HC$_3$N formation.
\section{A simple model of HC$_3$N shell}
From the discussion in the preceeding section we suggest that HC$_3$N emission traces the remnant AGB envelope around
the Egg nebula. The AGB envelope is disrupted by the interaction with the fast collimated
outflows. These outflows open channels or cavities along the polar and equatorial directions. 
Thus the HC$_3$N J=5--4 emission is very useful in providing complementary information to previous
observations on the structure of the Egg nebula. To gain more insight on the structure of the
nebula and to estimate the abundance of HC$_3$N, we construct a simple of model of the HC$_3$N shell. 
In Figure 4. we show a sketch of the overall 
structure of the HC$_3$N shell. The hollow shell is filled with the remnant AGB wind, which is
asummed to be spherically symmetric. Collimated fast outflows emanate from the central post-AGB star along both
polar and equatorial directions. These outflows interacts with the slow AGB wind, entraining
molecular gas along their path. As a result, the
remnant AGB envelope is disrupted and two pairs of cavities are excavated by the fast collimated outflow.
For simplicity, we assume that the abundance of HC$_3$N is constant over the whole hollow shell. We do not
take into account the presence of HC$_3$N J=5--4 emission near the center of the nebula because a correct 
derivation of the abundance and spatial distribution would require more constraints from multiline observations.
We also assume that the HC$_3$N J=5--4 line forms under the LTE condition
with a temperature of 30 K. Although the LTE assumption might be quite crude in the circumstellar
envelope environment, it is commonly used to estimate the relative abundance of molecules. For our purpose
of studying the spatial kinematics and estimate of the abundance, that assumption should be adaquate. When
multi-line observations are available, more elaborate model can be constructed. We project
a spherical hollow shell into a 3-dimension regular grid. We adopt an expansion velocity of 15 kms$^{-1}$
for the shell. The gas density at each radial point in the shell is calculated from the mass loss rate. 
We adopt the mass loss rate of 3x10$^{-5}$ M$_\odot$ yr$^{-1}$ estimated in the previous section. 
The turbulence velocity, which defines the local linewidth, is assumed to be 1 kms$^{-1}$
as is commonly used for circumstellar envelopes. 
The intensity of the HC$_3$N J=5--4 line is calculated by solving directly the radiative transfer equation along
each line of sight. We then use the emerging intensity to form the model channel maps. The channel maps are
then convolved with the synthesized beam of our VLA observation. The resulting model channel maps can be then
compared directly with our observation. We use a inclination angle of 15$^\circ$ for the envelope, which
is almost the same as derived from the model of Yusef-Zadeh et al. (1984). The opening angle of the cavities in the both the polar
direction and in the equatorial plane is 40$^\circ$. The position angle of the nebular axis on the plane of the
sky is set to 20$^\circ$, which is close to that advocated by Cox et al. (2000). The complex distribution of
HC$_3$N does not allow a more accurate estimate of the position angle. Besides, there are several outflows contributing
to the cavities in the polar direction, making the definition of the position angle somewhat ill-defined. 
The parameters of our simple model are collected in Table. 2.

In Figure 5 we show the channel maps of the HC$_3$N J=5--4 emission calculated with our model. 
The main features found in the observed channel maps, namely the cavities and the changing
morphology of the emission between velocity channels, are qualitatively reproduced in our model.
The predicted total intensity profile of the HC$_3$N J=5--4 line is shown in Figure 6.
The predicted line profile is similar to that
seen in the single dish observation of Fukasaku et al. (1994) and also in our observations. 
The observed strength (about 2 Jy at the peak intensity) and the parabolic shape with a depression 
around the systemic velocity, which is caused by the presence of the two pair of cavities, 
of this line is also reproduced with our model.  . 
\section{Summary}
We have imaged at high angular resolution the distribution of HC$_{\rm 3}$N J=5--4 emission line
in the Egg nebula  We find that in the approaching and receding portion of the envelope 
the HC$_{\rm 3}$N emission traces a  clumpy hollow shell structure, similar 
to that seen in normal carbon rich envelopes. Near the systemic velocity, however,
the hollow shell is fragmented into several large blobs or arcs. The missing portions of the hollow
shell correspond spatially to locations of the high velocity outlows observed previously in the
Egg nebula. We also find that the intersection of fast molecular outflows 
previously suggested as the location of the central post-AGB star is significantly offset from the
center of the hollow shell.
We interprete the observed spatial-kinematics of HC$_3$N J=5--4 emission as the 
direct evidence for the disruption of the slowly expanding  
envelope ejected during the AGB phase by the interaction with
collimated high velocity outflows initiated during the transition. 
From modelling the HC$_3$N distribution we could reproduce
qualitatively the spatial kinematics of the HC$_3$N J=5--4 emission
using a HC$_3$N shell with two pairs of cavities with opening angle of $\sim$40$^\circ$ 
opened by the collimated high velocity
outflows along the polar direction and in the equatorial plane.
We infer a relatively high abundance of HC$_3$N/H$_2$ 
$\sim$3x10$^{-6}$ for an estimated mass loss rate of 3 10$^{-5}$ M$_\odot$ yr$^{-1}$.
The high abundance of HC$_3$N and the presence of some weaker J=5--4 emission in the
vicinity of the central post-AGB star suggest an unusually efficient formation of this molecule in the Egg nebula.
\acknowledgments
We thank the VLA staff for their help with the observations. 
This research has made use of NASA's Astrophysics Data System Bibliographic Services
and the SIMBAD database, operated at CDS, Strasbourg, France.
\begin{table}
\caption{Summary of the observations}
\begin{tabular}{ll}\hline
Line               & HC$_3$N J=5--4 \\
Rest frequency     & 45.490316 GHz \\
Obs. date/Config.  & 24 Nov 2002 \\
                   & C configuration, 4 hours on-source time \\
                   & 10 Mar 2003 \\
		   & D configuration, 1 hour on-source time \\
Synthesized beam    &  1.3" x 1.08" at PA = 2$^\circ$.35 \\
$\Delta$V       &  3.9 kms$^{-1}$ \\
rms noise    & 2.3 mJy beam$^{-1}$ \\ 
\hline
\end{tabular}
\end{table}
\begin{table}[h]
\caption{Model parameters.}
\begin{tabular}{ll}
\hline
Distance            & 420 pc \\
Mass loss rate      & 3 10$^{-5}$ M$_\odot$ yr$^{-1}$ \\
Expansion velocity  & 17 kms$^{-1}$ \\
$[$HC$_3$N]/[H$_2$]   & 3 10$^{-6}$ \\
Inner radius of HC$_3$N shell  & 10$^{16}$ cm \\
Outer radius of HC$_3$N shell  & 4 10$^{16}$ cm \\
Turbulence velocity & 1 kms$^{-1}$ \\
Opening angle of bipolar lobes & 40$^\circ$ \\
Opening angle of equatorial cavities & 40$^\circ$ \\
Inclination angle   & 15$^\circ$ \\
Position angle of bipolar lobes  & 20$^\circ$ \\
\hline
\end{tabular}
\end{table}
\newpage

\begin{figure*}[ht]
\plotone{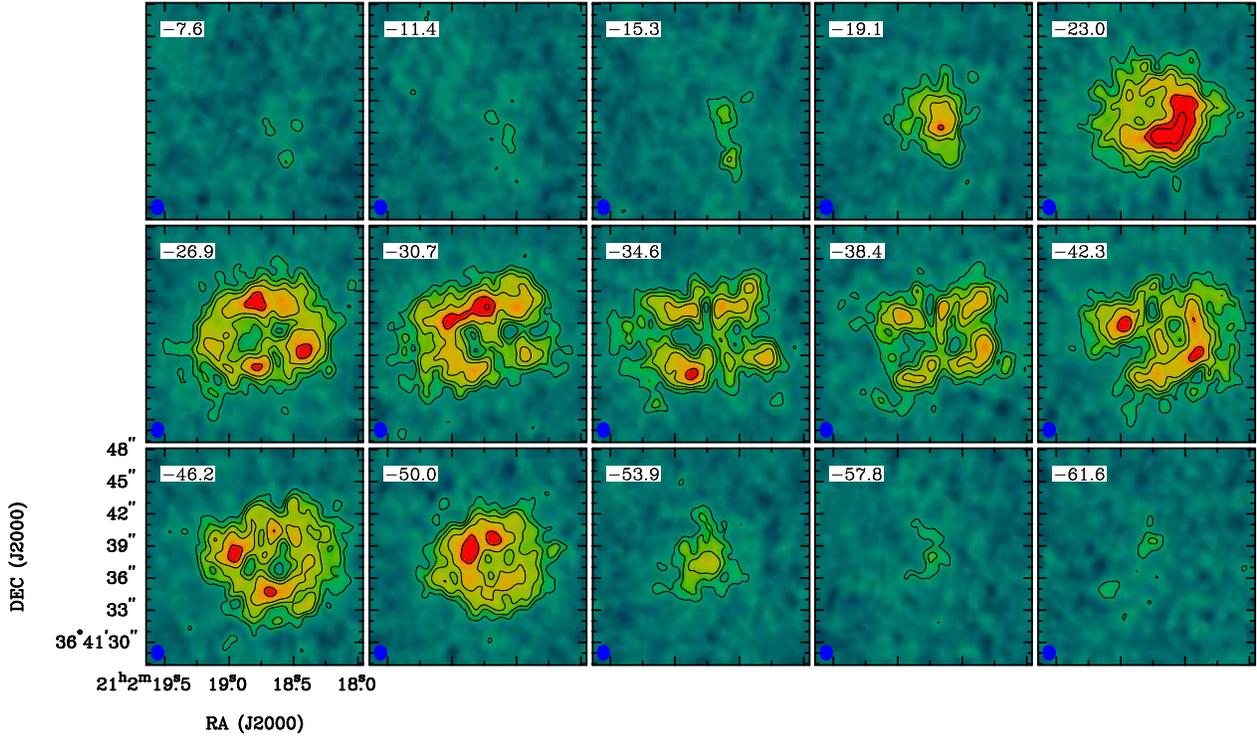}
\caption{Channel maps of the HC$_3$N J=$5-$4 emission from the Egg nebula in contour and grayscale. The contour levels
are (3, 5, 7, 10, 15, 20)$\sigma$ where the rms noise level $\sigma$ = 2.3 mJy beam$^{-1}$.
The synthesize beam of 1".3x1".08 is shown in the lower left of the upper left frame. The conversion factor
between the brightness temperature and the flux density of the HC$_3$N J=5--4 emission is 2.37 mJy K$^{-1}$.}
\label{fig1}
\end{figure*}

\begin{figure*}[ht]
\plotone{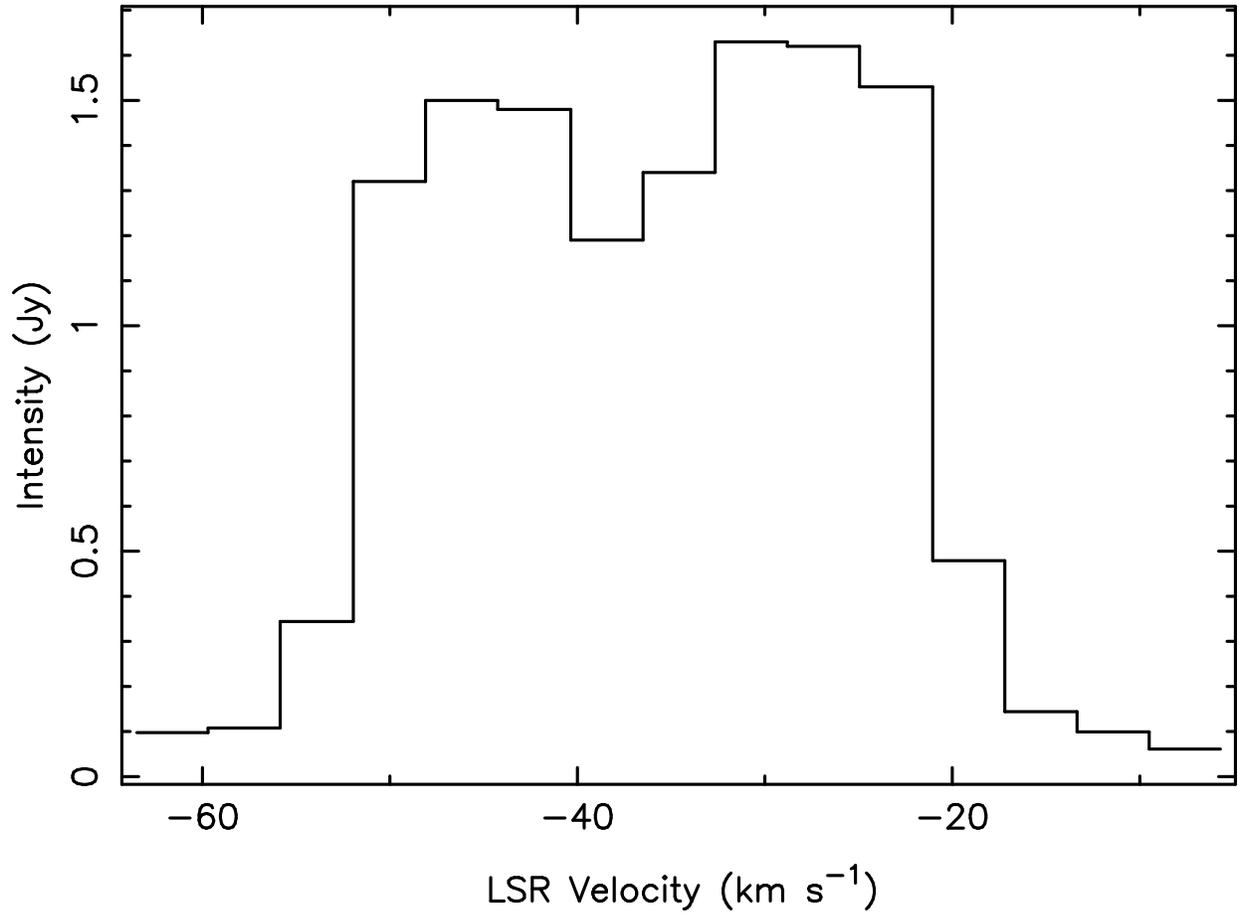}
\caption{Total intensity profile of the HC$_3$N J=$5-$4 line from the Egg nebula.}
\label{fig2}
\end{figure*}

\begin{figure*}[ht]
\plotone{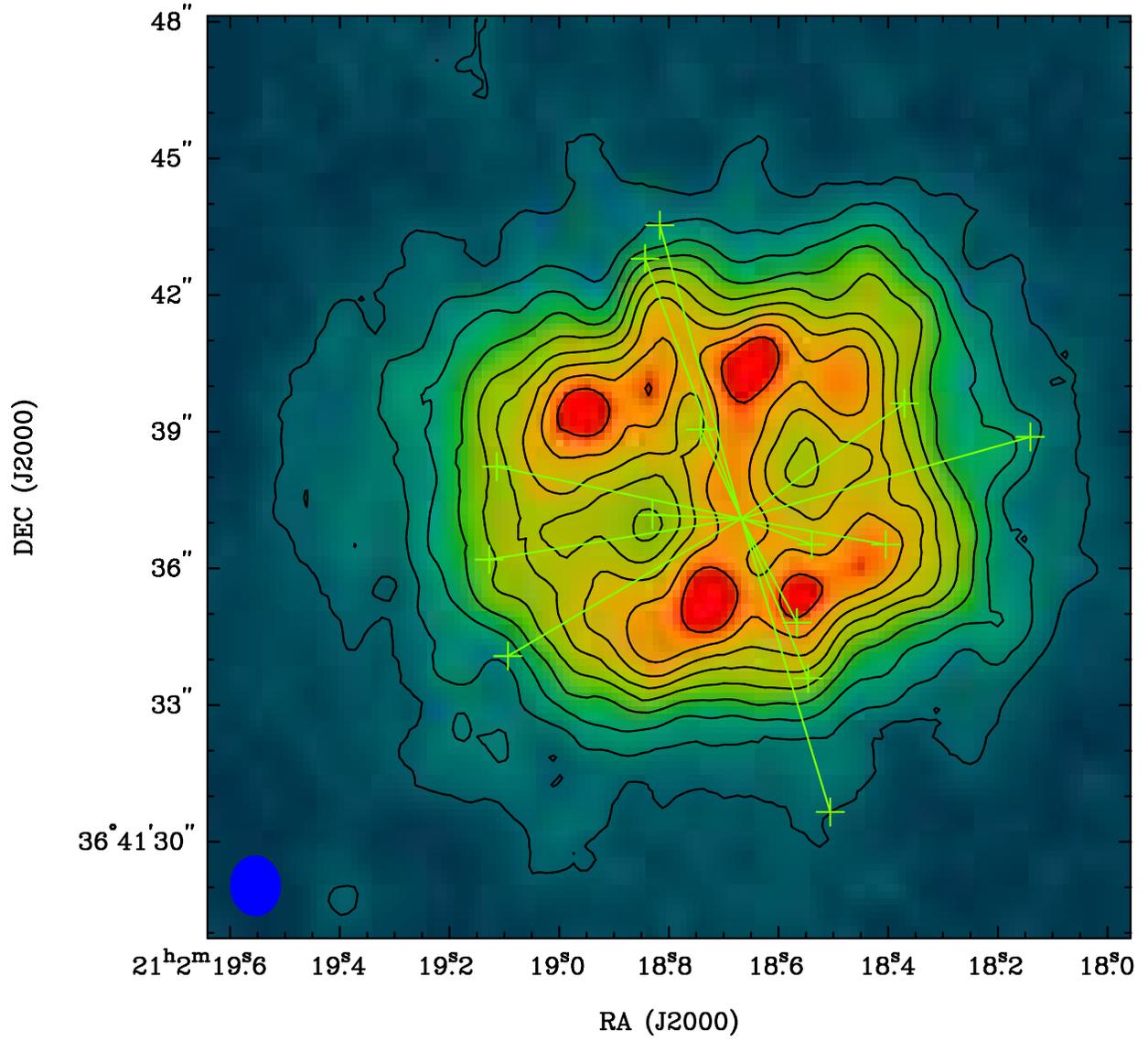}
\caption{The integrated intensity map of the HC$_3$N J=$5-$4 line from the Egg nebula. The contour
levels start from 0.1 Jy kms$^{-1}$ in step of 0.1 Jy kms$^{-1}$. The high velocity outflows identified
by Cox et al. (2000) are marked with croses and solid lines.}
\label{fig3}
\end{figure*}

\begin{figure*}[ht]
\plotone{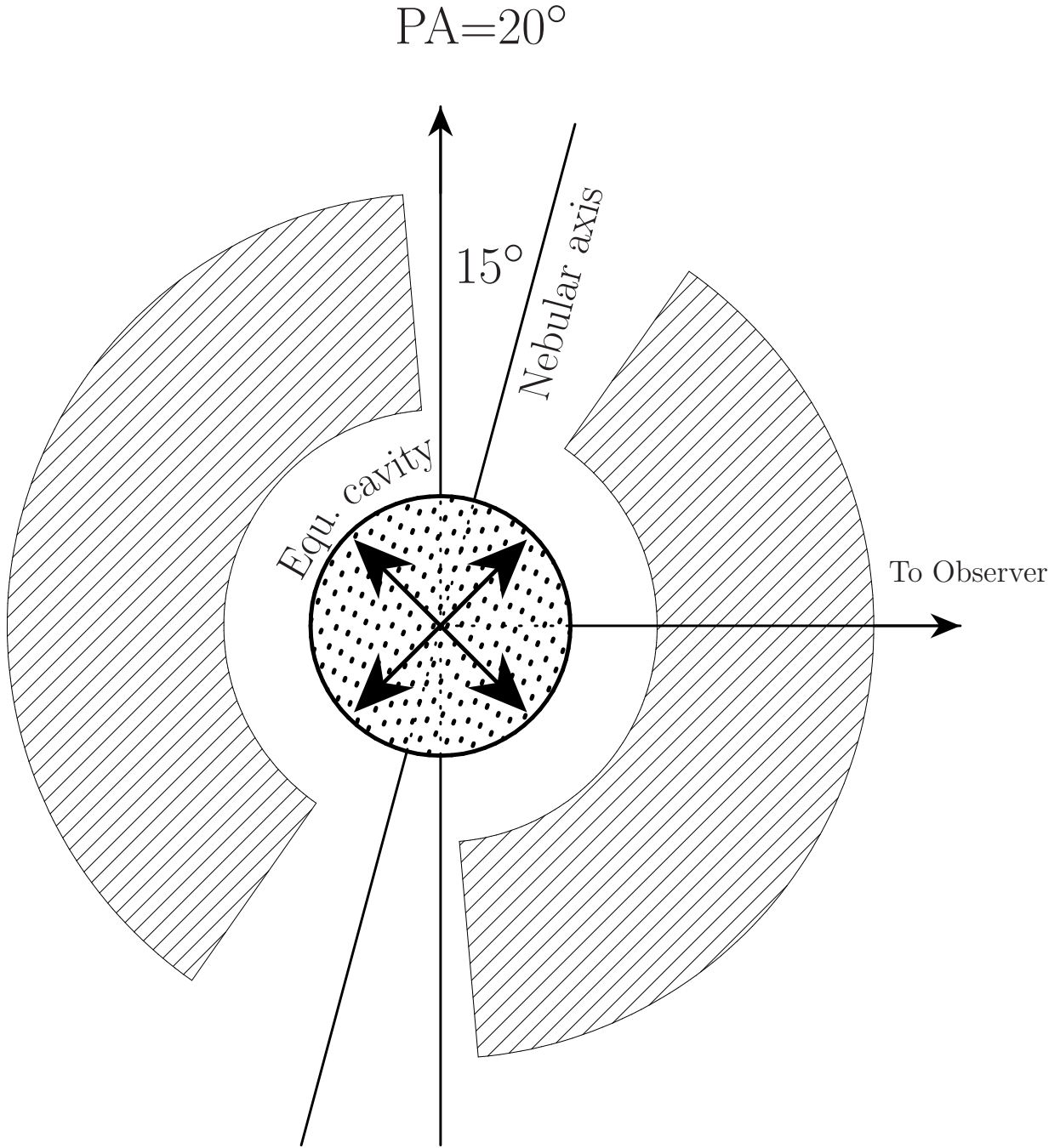}
\caption{Sketch of the HC$_3$N shell in the Egg nebula. The spherical shell has holes or cavities along the polar directions
and in the equatorial plane due to the disruptive effect of the high velocity jets.}
\label{fig4}
\end{figure*}

\begin{figure*}[ht]
\plotone{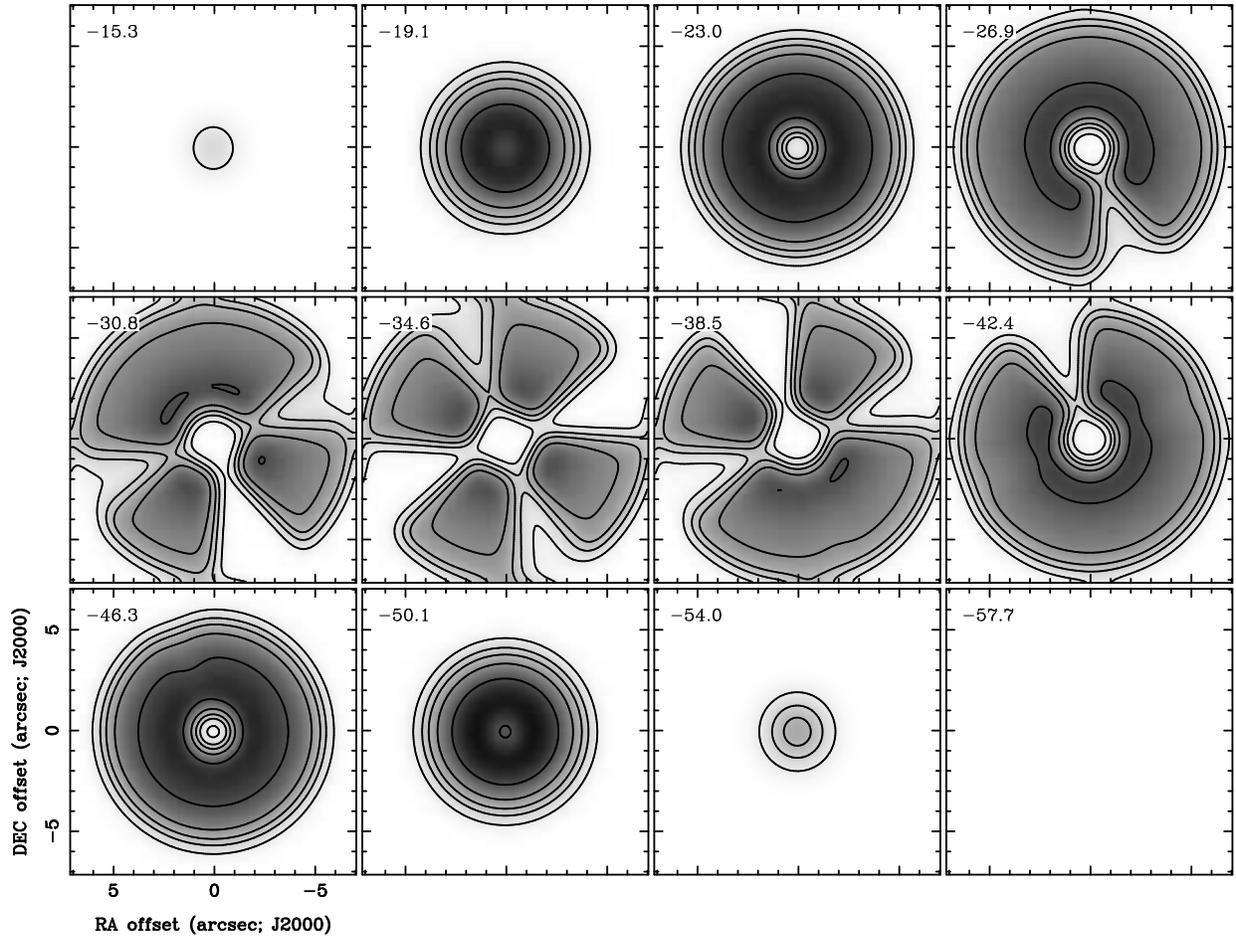}
\caption{Channel maps of HC$_3$N J=5--4 emission calculated with our model and convolved with the same gaussian
beam as the synthesized beam of our VLA observations. The emission is shown in
both greyscale and in contours. The contour levels are the same as in Figure 1.}
\label{fig5}
\end{figure*}

\begin{figure*}[ht]
\plotone{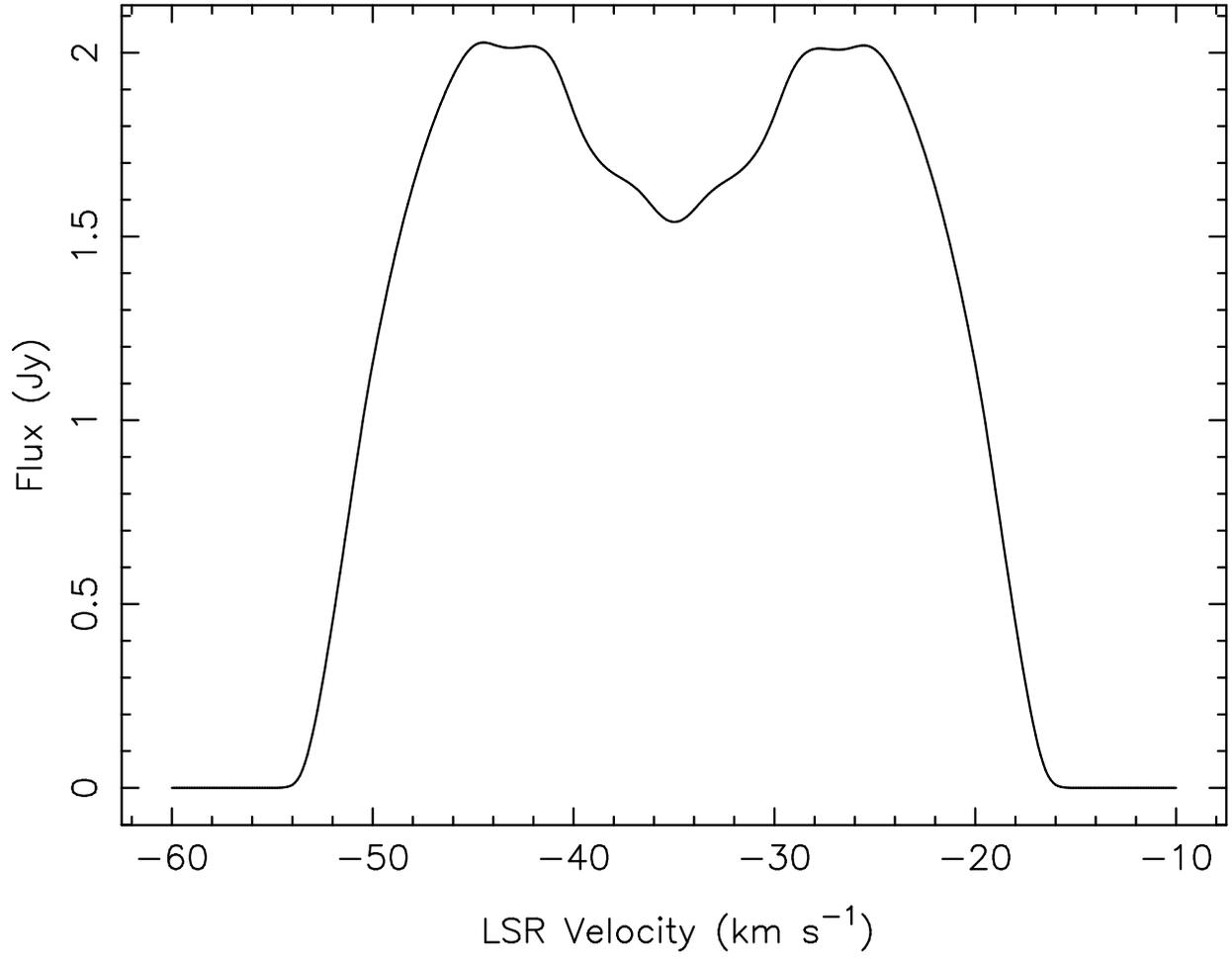}
\caption{Predicted total intensity profile of the HC$_3$N J=5--4 line.}
\label{fig6}
\end{figure*}


\begin{thebibliography}{}
\bibitem[Audinos et al. (1994)]{audinos94} Audinos, P., Kahane, C., Lucas, R., 1994, A\&A 287, L5
\bibitem[Bieging \& Tafalla (1993)]{bieging93} Bieging, J.H., Tafalla, M., 1993, AJ 105, 576
\bibitem[Bieging \& Nguyen-Q-Rieu (1996)]{bieging96} Bieging, J.H., Nguyen-Q-Rieu, 1996, AJ, 112, 706
\bibitem[Cherchneff et al. (1993)]{ch93} Cherchneff, I., Glassgold, A., Mamon, G., 1993, ApJ 410, 188
\bibitem[Chiu et al. (2006)]{chiu06} Chiu, P.J., Hoang, C.T., Dinh-V-Trung, et al., 2006, ApJ 645, 605
\bibitem[Cox et al. (2000)]{cox00}
Cox, P., Lucas, R., Huggins, P.J., Forveille, T., Bachiller, R., Guilloteau, S., 
Maillard, J.P., Omont, A., 2000, A\&A 353, L25
\bibitem[Crampton et al. (1975)]{crampton75}
Crampton, D., Cowley, A.P., Humphreys, R.M., 1975, ApJ, 198, L135
\bibitem[Crosas \& Menten (1997)]{crosas97} Crosas, M., Menten, K., 1997, ApJ 483, 913
\bibitem[Dinh-V-Trung \& Nguyen-Q-Rieu (2000)]{trung00} Dinh-V-Trung, Nguyen-Q-Rieu, 2000, A\&A 361, 601
\bibitem[Dinh-V-Trung \& Lim (2008)]{trung08} Dinh-V-Trung, Lim, J., 2008, ApJ 678, 303
\bibitem[Fukasaku et al. (1994)]{faka94} Fakasaku, S., Hirahara, Y., Masuda, A., et al., 1994, ApJ 437, 410
\bibitem[Jura et al. (2001)]{jura01} Jura, M., Turner, J.L., van Dyk, S., Knapp, G.R., 2001, ApJ 528, L105
\bibitem[Klochkova et al. (2000)]{klochkova00}
Klochkova, V.G., Szczerba, R., Panchuck, V.E., 2000, Astr. Lett., 26, 439
\bibitem[Knapp et al. (1993)]{knapp93}
Knapp, G.R., Sandell, G., Robson, E.I., 1993, \apjs, 88, 173
\bibitem[Latter et al. (1993)]{latter93}
Latter, W.B., Hora, J.L., Kelly, D.M., Duetsch, L.K., Maloney, P.R., 1993, AJ, 106, 260
\bibitem[Lee \& Sahai (2003)]{lee03} Lee, C.F., Sahai, R., 2003, ApJ 586, 319
\bibitem[Lovas (2004)]{lovas04} Lovas, F.J., 2004, J.Phys.Chem.Ref.Data, 33, 177
\bibitem[Lucas \& Gu\'{e}lin (1999)]{lucas99} Lucas, R., Gu\'{e}lin, M., 1999, IAU Symposium 191, edited
by T. Le Bertre, A. Lebre and C. Waelkens, p. 305 
\bibitem[Millar et al. (2000)]{millar00} Millar, T.J., Herbst, E., Bettens, R.P.A., 2000, MNRAS 316, 195
\bibitem[Ney et al. (1975)]{ney75}
Ney, E.P., Merrill, K.M., Becklin, E.E., Neugebauer, G., Wynn-Williams, C.G., 1975, ApJ 198, L129
\bibitem[Nguyen-Q-Rieu \& Bieging (1990)]{rieu90} Nguyen-Q-Rieu, Bieging, J., 1990, ApJ 359, 131
\bibitem[Sahai et al. (1998a)]{sahai98a} Sahai, R., Trauger, J.T., Watson, A.M., et al. 1998, ApJ 493, 301
\bibitem[Sahai et al. (1998b)]{sahai98b} Sahai, R., Hines, D., Kastner, J.H., et al., 1998, ApJ 492, L163
\bibitem[Shoier et al. (2002)]{schoeir02} Schoier, F., Ryde, N., Olofsson, H., 2002, A\&A 391, 577
\bibitem[Truong-Bach et al. (1993)]{bach93} Truong-Bach, Graham, D., Nguyen-Q-Rieu, 1993, A\&A 277, 133
\bibitem[Ueta et al. (2006)]{ueta06}
Ueta, T., Murakawa, K., Meixner, M., 2006, ApJ, 641, 1113
\bibitem[Yamamura et al. (1996)]{yamamura96}
Yamamura, I., Onaka, T., Kamijo, F., Deguchi, S., Ukita, N., 1996, ApJ, 465, 926
\bibitem[Yusef-Zadeh et al. (1984)]{yusef84} Yusef-Zedeh, F., Morris, M., White, R.L., 1984, ApJ 278, 186
\end{thebibliography}
\end{document}